\documentclass[pre,aps,
twocolumn,pdflatex,
superscriptaddress,floats,showpacs,10pt]{revtex4-1}
\usepackage{graphicx}
\usepackage{dcolumn}
\usepackage{bm}
\usepackage{amssymb}
\usepackage{amsmath}
\usepackage{array}
\usepackage{float}
\usepackage{geometry}
\usepackage{color}
\usepackage{amsthm}

\begin{document}

\title{On extreme field limits in high power laser matter interactions: 
radiation dominant regimes in high intensity electromagnetic wave interaction with electrons}
\author{ Sergei V. Bulanov$^{a,b,c}$, Timur Zh. Esirkepov$^{a}$, Masaki Kando$^{a}$,\\ 
James K. Koga$^{a}$, Tatsufumi Nakamura$^{a}$, Stepan S. Bulanov$^{d}$, \\ 
Alexei G. Zhidkov$^{e}$, Yoshiaki Kato$^{f}$, Georg Korn$^{g}$\\
{\small $^{a}$Kansai Photon Science Institute, JAEA, Kizugawa, Kyoto 619-0215, Japan}\\
{\small $^{b}$Prokhorov Institute of General Physics, RAS, Moscow 119991, Russia}\\
{\small $^{c}$Moscow Institute of Physics and Technology, Dolgoprudny, }\\
{\small Moscow region 141700, Russia}\\
{\small $^{d}$University of California, Berkeley, CA 94720, USA}\\
{\small $^{e}$Photon Pioneers Center, Osaka University, Suita, Osaka 565-0871, Japan}\\
{\small $^{f}$The Graduate School for the Creation of New Photonics Industries, Japan}\\
{\small $^{g}$ELI-Beamlines, Institute of Physics, CAS, Prague, Czech Republic}
}
\maketitle 
  
{\bf Abstract}
We discuss the key important regimes of electromagnetic field interaction with charged particles. 
Main attention is paid to the nonlinear Thomson/Compton scattering regime with the radiation friction and 
quantum electrodynamics effects taken into account. This process opens a channel 
of high efficiency electromagnetic energy conversion into 
hard electromagnetic radiation in the form of ultra short high power gamma ray flashes.



\section{Introduction}

Radiation of present day lasers approaches the intensity regimes where in the electromagnetic 
(EM) wave interaction with matter 
the radiation friction effects on the charged particle dynamics become dominant \cite{MTB, DPHK}. 
At these limits the electron dynamics become dissipative \cite{YaBZ, ZKS, KEB, Thomas} with
fast conversion of the EM wave energy to hard EM radiation, which for
typical laser parameters is in the gamma-ray range \cite{Ridgers, NakaKo}.
For laser radiation with $1\,\mu$m wavelength the radiation friction force changes the scenario
of the EM wave interaction with matter at the intensity of about $I_R\approx 10^{23}$W/cm$^{2}$. 
For the laser intensity close to $I_R$ also novel physics of abundant electron-positron
pair creation comes into play \cite{BELL, KIRK} (see \cite{Fedotov, Nerush} 
and \cite{SSB, NIMA}). In this regime, the electron (positron) interaction with the EM field is
principally determined by a counterplay between the radiation friction and quantum effects. 
The quantum electrodynamics (QED) effects weaken the EM emission by the relativistic electron resulting in 
the lowering of the radiation friction \cite{JSCHW, AAS}. In an extremely high intensity EM field 
vacuum looses its property to be a empty substance showng such various nonlinear QED processes  as 
vacuum polarization, electron-positron pair plasma creation, and other properties depending on the EM field 
amplitude. 

\begin{figure}[tbph]
\centering
\includegraphics[width=7cm,height=4.8cm]{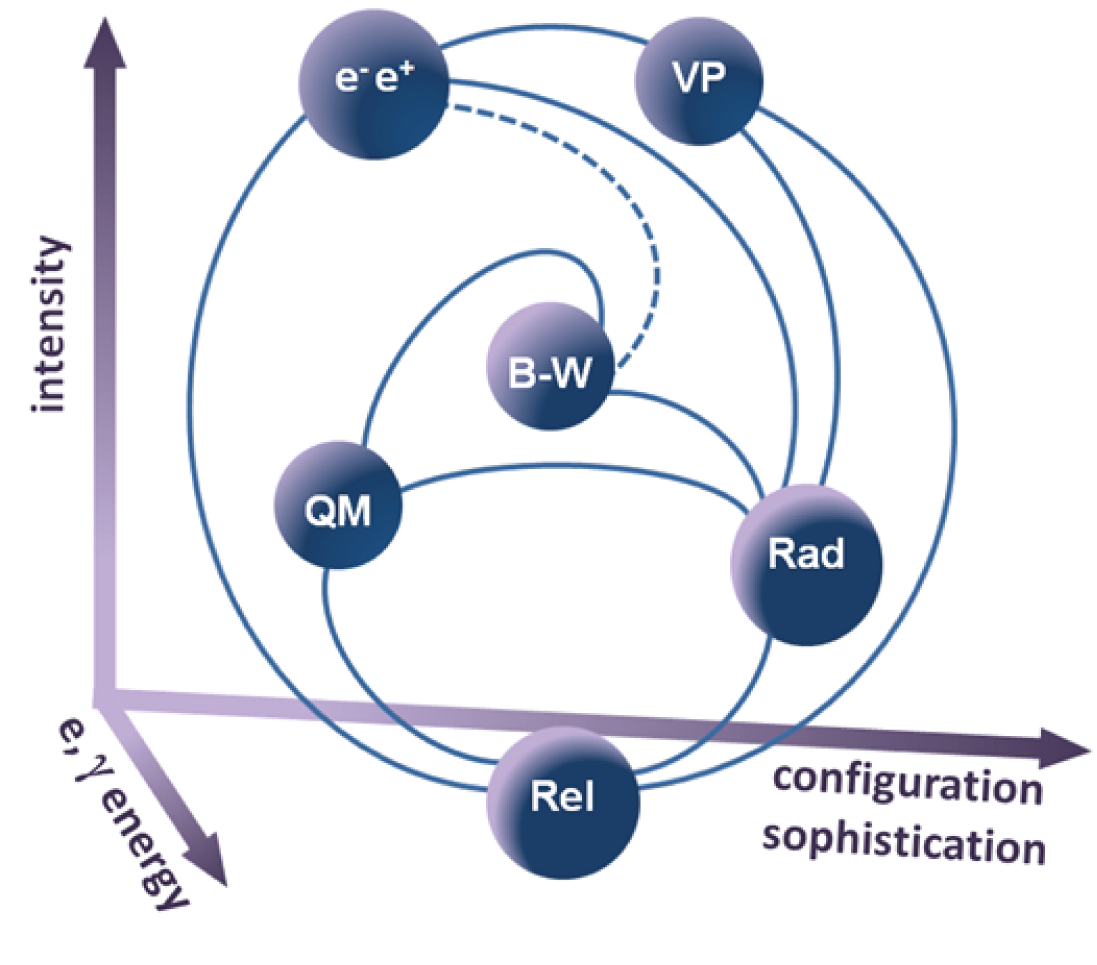}
 \caption{ Schematic of nonlinear QED processes. 
Laser-matter interaction transition from the relativistic regime (Rel) to dominant radiation friction (RF) and 
nonlinear Thomson scattering (NTS) through the limit when quantum electrodynamics comes into play (QED) and to 
 electron-positron avalance/cascade development (A/C), towards nonlinear vacuum with electron-positron creation (${\rm E_S}$) and 
nonlinear vacuum polarization (VP) while the EM field intensity increases.}
\label{FIG1}
\end{figure}

In Fig. \ref{FIG1} we present a schematic 
of nonlinear QED processes which realization depends not only on the EM radiation intensity, 
but also on the photon and charged particle energy and on the EM field configuration \cite{SSB, Mult, DAS}. 
Fig. \ref{FIG1} illustrates a transition from the relativistic interaction regime (Rel) to dominant radiation friction (RF) and 
nonlinear Thomson scattering (NTS) through the limit when the quantum electrodynamics comes into play (QED) and to 
 electron-positron avalance/cascade development (A/C), towards nonlinear vacuum with electron-positron creation (${\rm E_S}$) and 
nonlinear vacuum polarization (VP) while the EM field intensity grows.

The probabilities of the processes involving extremely high intensity EM field interaction with 
electrons, positrons and photons are determined by several dimensionless parameters.

When the normalized dimensionless EM wave amplitude
\begin{equation}
a=\frac{e E \lambda}{2 \pi m_e c^2}
\label{eq:parameters}
\end{equation}
exceeds unity, $a>1$, the energy of the electron quivering in the field of the wave becomes relativistic.
Here $\lambda=2 \pi c/\omega$ with $\omega$ being the EM wave frequency. 
The EM wave intensity is expressed via the normalized amplitude, $a$, as
\begin{eqnarray}
I_{rel}=\frac{m_e^2 c^3 \omega^2 a^2}{4 \pi e^2} \nonumber \\
=1.37\times 10^{18} a^2 \left(\frac{1 \mu {\rm m}}{\lambda}\right)^2\frac{\rm  W}{{\rm cm}^2}.
\label{eq:Irel}
\end{eqnarray}
In a plane EM wave the parameter $a$ is related to the Lorentz invariant, which being expressed via
the 4-potential of the electromagnetic field, $A^{\mu}$, is equal to $a=e\sqrt{A^{\mu}A_{\mu}}/m_e c^2$.
Here and below a summation over repeating indices $\mu=0,1,2,3$ is assumed.

A relativistic electron interacting with an EM wave emits high energy photons. 
Here and below for the sake of simplicity we analyse the dynamics of a radiating electron in a homogeneous rotating electric field,
which corresponds to the nodes of  two colliding EM waves, where the wave magnetic field vanishes, 
and/or to the electron interaction with an EM wave in near-critical plasmas
in the frame of reference moving with the group velocity of the wave \cite{BOOST}.

In the regime of Nonlinear Thomson Scattering (NTS) 
the power emitted 
is proportional to the fourth power of its energy, $m_e c^2 \gamma$, \cite{LL-TP}
\begin{equation}
P_{\gamma}\approx\varepsilon_{rad} m_e c^2 \omega \gamma_e^4.
\label{eq:NTS1}
\end{equation}
The dimensionless parameter, 
\begin{equation}
\varepsilon_{rad}=\frac{4 \pi r_e}{3 \lambda}=1.17\times 10^{-8} \left(\frac{1 \mu {\rm m}}{\lambda}\right),
\label{eq:eprad}
\end{equation}
proportional to the ratio of the classical electron radius $r_e=e^2/m_e c^2=2.8\times 10^{-13}$ cm
and the EM wave wavelength $\lambda$ characterizes the role of radiation losses. 
The maximal rate at which an electron can acquire
the energy from the EM field is approximately equal to $m_e c^2 \omega a$.
The condition of the balance between the acquired and
lost energy for the electron Lorentz factor equal to $\gamma_e=a$ shows that the
radiation effects become dominant at $a>a_{rad}=\varepsilon_{rad}^{-1/3}$, i.e. at 
the EM wave intensity above 
\begin{eqnarray}
I_R=\left(\frac{3}{2}\right)^{2/3}\frac{m_e^{8/3} c^5 \omega^{4/3}}{4 \pi e^{10/3}} \nonumber \\
=2.65 \times 10^{23} \left(\frac{1 \mu {\rm m}}{\lambda}\right)^{4/3}\frac{\rm W}{{\rm cm}^2}.
\label{eq:Irad}
\end{eqnarray}

The characteristic frequency of the emitted radiation is proportional to the cube of the electron energy,
$\omega_m\approx 0.3 \omega \gamma_e^3$. QED effects become important, when
the energy of the photon generated by Thomson (Compton) scattering is of
the order of the electron energy, i.e. $\hbar \omega_m \approx m_e c^2 \gamma_e$. 
If $\gamma_e=a$ this yields the QED limit on the EM field amplitude, $a^2/a_S>1$. Here 
the dimensionless parameter
\begin{eqnarray}
a_S=\frac{e E_S \lambda}{2 \pi m_e c^2}=\frac{m_e c^2}{\hbar \omega} \nonumber \\
=\frac{\lambda}{\lambda_C}=4.2\times 10^{5}\left(\frac{\lambda}{1 \mu {\rm m}}\right)
\label{eq:aS}
\end{eqnarray}
is the normalized critical electric field of quantum electrodynamics, 
$E_S=m_e^2 c^3/e \hbar$ \cite{BLP}, with $\lambda_C=2 \pi \hbar /m_e c=2.42\times 10^{-10}$cm 
being the Compton wavelength. The EM radiation intensity for the wave with the amplitude $a_S$ is 
\begin{equation}
I_S=\frac{m_e^4 c^7}{4 \pi e^2 \hbar^2}=2.36 \times 10^{29} 
\frac{\rm  W}{{\rm cm}^2}.
\label{eq:IS}
\end{equation}
The above obtained  QED limit, $a^2/a_S>1$, corresponds to the condition $\chi_e >1$, where relativistic and gauge invariant parameter $\chi_e$
\begin{equation}
\chi_e=\frac{\sqrt{\left(F^{\mu \nu} p_{\nu}\right)^2}}{E_S m_e c }\approx 2\frac{a}{a_S}\gamma_e
\label{eq:chie}
\end{equation}
characterizes the probability of the gamma-photon emission by the electron with 4-momentum $p_{\nu}$ in the field of the EM wave. 
The 4-tensor of the EM field is defined as $F_{\mu \nu}=\partial_{\mu} A_{\nu}-\partial_{\nu} A_{\mu}$. The QED limit is reached for 
the EM wave intensity of the order of 
\begin{equation}
I_Q=\frac{m_e^3 c^5 \omega}{8 \pi e^2 \hbar}=5.75 \times 10^{23} 
\left(
\frac{1 \mu {\rm m}}
{\lambda}
\right)
\frac{\rm  W}{{\rm cm}^2}.
\label{eq:IQ}
\end{equation}
\begin{figure}[tbph]
\includegraphics[width=7cm,height=4.8cm]{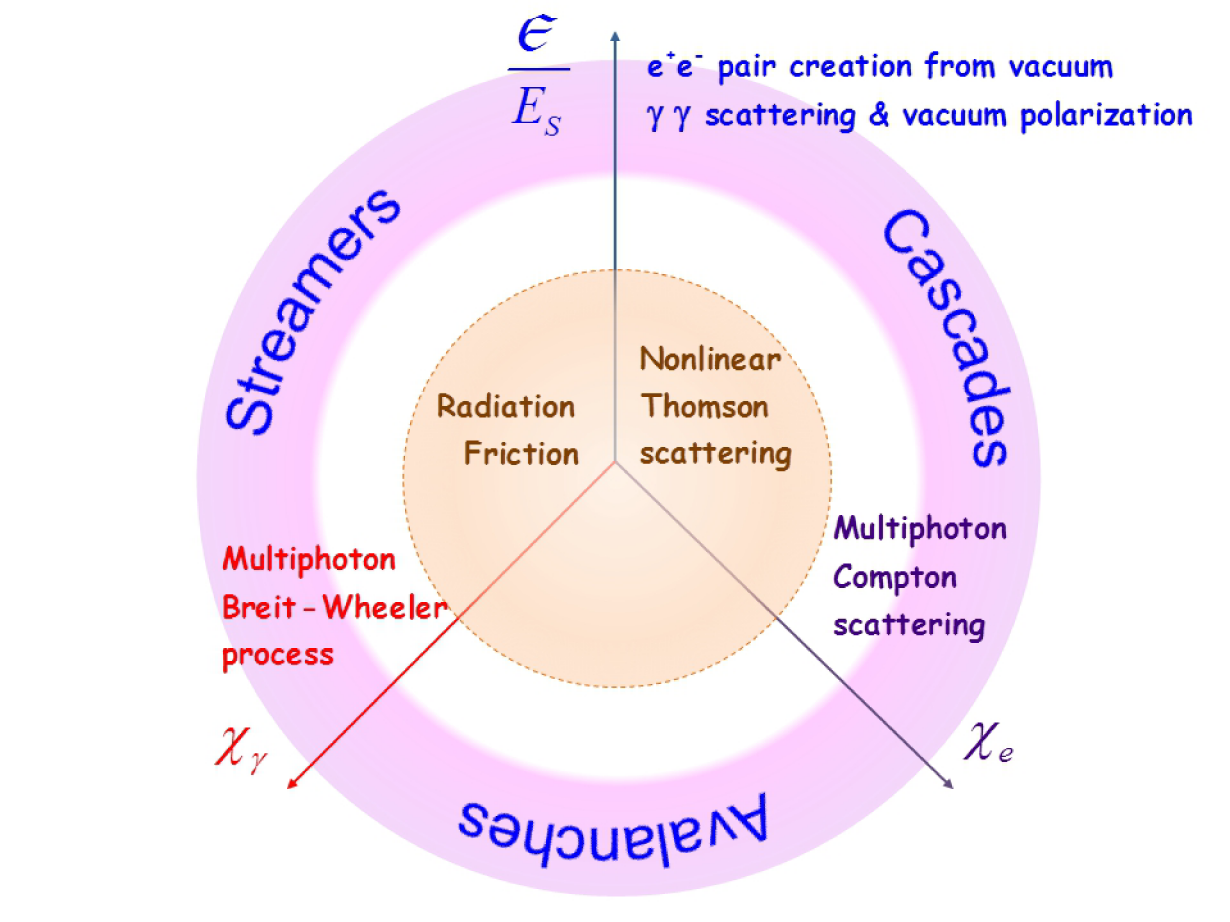}
\caption{ Regimes of EM field interaction in the space of parameters ${\mathbb E}/E_S$, $\chi_{\gamma}$, and $\chi_e$. 
The limit ${\mathbb E}/E_S \to 1$ corresponds to the nonlinear QED vacuum regimes with the electron-positron pair 
ceation from vacuum and photon-photon interaction.
When the parameter $\chi_e$ becomes large, multiphoton Compton scattering results in the high energy photon 
generation. For $\chi_{\gamma} > 1$ the mutiphoton
Breit-Wheeler process results in electron-positron pair generation via the gamma-photon interaction 
with a strong EM field. The nonlinear Thomson scattering regime 
is realized for $a \gg 1$ with the radiation friction effects coming in play at $a>a_{rad}$. \label{FIG2}}
\end{figure}

The dimensionless parameter 
\begin{equation}
\chi_{\gamma}=\frac{\hbar \sqrt{\left(F^{\mu \nu} k_{\nu}\right)^2}}{E_S m_e c }\approx 2\frac{a}{a_S^2}\frac{\omega_{\gamma}}{\omega}
\label{eq:chig}
\end{equation}
determines the probability of the electron-positron pair creation by the photon with the energy $\hbar \omega_{\gamma}$ in the EM field via the
Breit-Wheeler process \cite{BW, NR}.

Nonlinear vacuum properties are determined by the Poincare invariants, 
\begin{equation}
{\mathbb F}=({\bf B}^2-{\bf E}^2)/2 
\end{equation}
and
\begin{equation}
{\mathbb G}=({\bf E \cdot B}). 
\end{equation}
The probability 
of the electron-positron pair creation depends on 
\begin{equation}
{\mathbb E}=\sqrt{\sqrt{{\mathbb F}^2+{\mathbb G}^2}-{\mathbb F}} 
\end{equation}
and
\begin{equation}
{\mathbb B}=\sqrt{\sqrt{{\mathbb F}^2+{\mathbb G}^2}+{\mathbb F}} ,
\end{equation}
 which are the electric and magnetic fields in the frame of reference where they are 
parallel.

Figure \ref{FIG2} illustrates different regimes of the EM interaction in the space of parameters ${\mathbb E}/E_S$, $\chi_{\gamma}$, and $\chi_e$. 
In the limit ${\mathbb E}/E_S \to 1$ 
the EM waves can create electron-positron pairs from vacuum \cite{DPHK, VR}, the EM wave can interact via photon-photon collisions \cite{JKK} 
and vacuum polarization \cite{NNR}. When the parameter $\chi_e$ becomes large the multiphoton Compton scattering results in the high 
energy photon generation. For $\chi_{\gamma} > 1$ the mutiphoton
Breit-Wheeler process results in electron-positron pair generation via the gamma-photon interaction with the strong EM field. 
The nonlinear Thomson scattering (NTS) regime 
is realized for $a\gg 1$ with the scattering cross section depending on the EM field amplitude. 
If  $a>a_{rad}=\varepsilon^{-1/3}_{rad}$ radiation friction effects play a key role.

The comparision of expressions given by Eqs. (\ref{eq:Irad}) and (\ref{eq:IQ}) shows 
that for the EM wave length equal to $\lambda\approx 0.8 \mu {\rm m}$ the intensities 
$I_R$ and $I_Q$ are of the same order of magnitude as has been noted in Ref. \cite{BRADY}. 
The curves $I_R(\omega)$ and $I_Q(\omega)$ intersect to each other at the frequency equal to 
\begin{equation}
\omega_1=\frac{e^4 m_e}{18 \hbar^3}
\label{eq:om1}
\end{equation}
corresponding to the wavelength of $\lambda_1=0.821 \mu$m and the photon energy of the order of 1.5 eV.

It is convenient to rewrite expressions (\ref{eq:Irad}) and (\ref{eq:IQ}) for $I_R(\omega)$ and $I_Q(\omega)$ in terms of the frequency $\omega_1$ as
\begin{eqnarray}
{I_R=\frac{m_e^4 c^5 e^2}{144 \pi \hbar^4}\left(\frac{\omega}{\omega_1}\right)^{4/3}} \nonumber \\
{
=3.8 \times 10^{23} \left(\frac{821{\rm nm}}{\lambda}\right)^{4/3}\frac{\rm W}{{\rm cm}^2}}
\label{eq:Irad1}
\end{eqnarray}
and 
\begin{eqnarray}
I_Q=\frac{m_e^4 c^5 e^2}{144 \pi \hbar^4}\left(\frac{\omega}{\omega_1}\right)  \nonumber \\
=3.8 \times 10^{23} \left(\frac{821{\rm nm}}{\lambda}\right)\frac{\rm W}{{\rm cm}^2}.
\label{eq:IQ1}
\end{eqnarray}

In the present paper we mainly pay attention to the nonlinear Thomson/Compton scattering regime when the radiation friction and 
quantum electrodynamics effects play comparibly important roles.

	 \section{Radiation Friction Effects on Charged Particle Motion}

	 In order to describe the relativistic electron dynamics in the EM field we shall use 
	 the equations of electron motion with the radiation friction force 
	 in the Landau-Lifshitz form \cite{LL-TP} with a form-factor taking into account the QED weakening of the radiation friction. 
	The EM wave is modeled by a rotating electric field, which as noted above  corresponds to the transformation to 
	the boosted frame of reference moving with the group velocity of the wave. In this frame of reference 
	the electron 
	 equations of motion can be written as
\begin{eqnarray}
	 \frac{d {\bf q}}{d \tau}=
	 -{\bf  a} 
	 -\frac{\varepsilon_{rad} \, G_e(\chi_e)}{\gamma_e}
	 \left\{
	 \gamma_e^2 \frac{{d \bf { a}}}{d \tau} \right . 
	 \nonumber \\
	 \left . -{\bf { a}}
	 \left({\bf q}\cdot {\bf { a}}\right)+{\bf q} \left[\left(\gamma_e {\bf { a}}\right)^2
	 -\left({\bf q}\cdot {\bf { a}}\right)^2\right]
	 \right\},
	\label{EquMot}
\end{eqnarray}
where $\gamma=\sqrt{1+q_1^2+q_2^2+q_3^2}$ is the electron Lorentz factor. 
The form factor $G_e(\chi_e)$ with $\chi_e$ defined by Eq. (\ref{eq:chie}) describes the radiation friction reduction due to
 quantum effects. 	In 3D notation the parameter $\chi_e$ given by Eq. (\ref{eq:chie}) reads
	\begin{equation}
	 \chi_e=\frac{\gamma_e}{E_S}\sqrt{\left({\bf E}+\frac{1}{c}{\bf v}\times {\bf B}\right)^2-\left({\bf v}\cdot {\bf E}\right)^2}.
	\label{chi-e}
	\end{equation}
 
We 
introduce the normalized variables, 
     \begin{equation}
	 \tau=\omega t, \quad {\bf q}=\frac{{\bf p}}{m_e c}, \quad {\bf a}=\frac{e{\bf E}}{m_e \omega c}.
	\label{NormVar}
	\end{equation}
The dimensionless parameter defined by Eq. (\ref{eq:eprad}) determines the role of the radiation friction.

	As well known, QED effects weaken the radiation friction because in the quantum regime the charged particle emits less radiation.
	In the QED regime the recoil due to photon emission becomes important. 
	According to Refs. \cite{JSCHW} the total radiated intensity is reduced by a factor depending on the quantum parameter $\chi_e$. 
	In addition, in Eq. (\ref{EquMot}) following an approach used in Refs. \cite{BELL, Ridgers, BKS, HIP2} we take into account the QED effects by using the form-factor, 
	$G(\chi_e)$, equal to the ratio
	of the full radiation intensity, $I$, to the intensity emitted by a classical electron, $I_{cl}$. Using the results of Refs. \cite{RITUS, BLP, IVS}
	we can write the form-factor $G(\chi_e)$ as
	$$
	G_e(\chi_e)=
	$$
	\begin{equation}
	\int^{\infty}_0
	\frac{12+15 \chi_e x^{3/2}+12 \chi_e^2 x^3}{4\left(1+\chi_e x^{3/2}\right)^4}
	\Phi^{\prime}(x) x dx,\,\,\,\,
	\label{G-chi-e}
	\end{equation}
	where $\Phi(x)$ is the Airy function \cite{AS}.  Here, we neglect the effects of the discret nature of the photon emission 
	in the QED (see \cite{DUCLOUS, BRADY, HIP2}).
	
	In the limit $\chi_e\ll1$ the form-factor $G(\chi_e)$
	tends to unity as 
	$$
	G_e(\chi_e)=
		1-\frac{55 \sqrt{3}}{16}\chi_e+48 \chi_e^2- ... 
	$$
	\begin{equation}	
	\approx 1-5.95 \chi_e+48 \chi_e^2-...\,.\qquad \qquad \qquad 
	\label{G-chi-e0}
		\end{equation}
	For $\chi_e\gg1$ it tends to zero as
	$$
	G_e(\chi_e)=
	\frac{32 \pi}{27\, 3^{5/6} \Gamma (1/3)\chi_e^{4/3}}-\frac{1}{\chi_e^{2}}
	$$
	$$
	+\frac{110 \pi}{81\, 3^{1/6} \Gamma (2/3)\chi_e^{8/3}}-\frac{11\,3^{1/2}}{5\chi_e^{3}}+... 
	$$
	\begin{equation}
	\approx \frac{0.5564}{ \chi_e^{4/3}}-\frac{1}{\chi_e^{2}}
	+\frac{2.6 }{\chi_e^{8/3}}-\frac{3.81}{\chi_e^{3}}+...\,.
		\label{G-chi-eInf}
	\end{equation}

\begin{figure}[tbph]
\centering
\includegraphics[width=7cm,height=4.8cm]{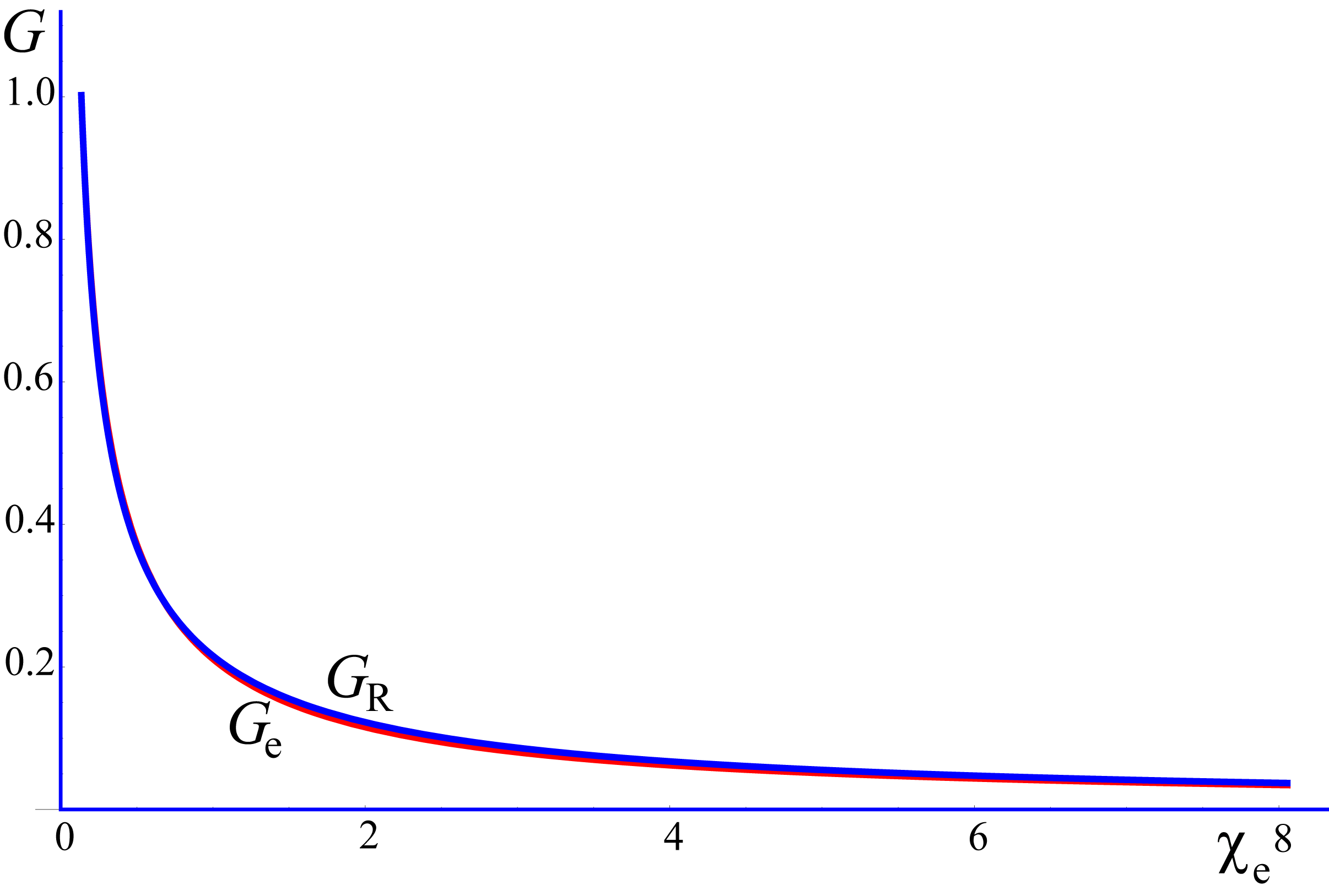}
\caption{Functions $G_e $ (lower curve) and $G_R $ (upper curve) given by Eqs. (\ref{G-chi-e}) and (\ref{G-chi-eappr}), respectively, versus $\chi_e$.
\label{FIG3}
}
\end{figure}

	However expression (\ref{G-chi-e}) and asymptotical dependences (\ref{G-chi-e0}) and (\ref{G-chi-eInf}) 
	are not convenient for implementing them in computer codes.
    For the sake of calculation simplicity  the approximation 	
	$$
	G_{BKS}(\chi_e)\approx 
	$$
	\begin{equation}
	\frac{1}{\left[1+4.8(1 + \chi_e) {\rm ln}(1 + 1.7\chi_e) + 2.44\chi_e^2\right]^{2/3}}.
	\label{G-chi-eapprBK}
	\end{equation}
	is usually used \cite{BKS}. This expression can be further simplified. The function
	\begin{equation}
	G_{R}(\chi_e)\approx \frac{1}{\left(1+8.93\chi_e + 2.41\chi_e^2\right)^{2/3}},
	\label{G-chi-eappr}
	\end{equation}
	which we shall use below, within the interval $0<\chi_e<10$ has an accuracy of approximation better than 1$\%$. 
	In Fig. \ref{FIG3} we plot the functions $G_e (\chi_e)$ and $G_R (\chi_e)$ given by Eqs. (\ref{G-chi-e}) and (\ref{G-chi-eappr}), respectively.
	We see that that their difference is negligebly small.
	
We consider a rotating electric field given by 
\begin{equation}
{\bf a}=
-a \left( 
{\bf e}_2 \cos \tau + {\bf e}_3 \sin \tau 
\right),
\end{equation}
where ${\bf e}_2$ and ${\bf e}_3$ are unit vectors along the coordinate axis in the plane 
perpendicular to the direction of the EM wave propagation.
We represent the electron momentum as 
(see also Ref. \cite{LADvsLL})
	\begin{equation}
\left( 
\begin{array}{c}
q_{1} \\ 
q_{||} \\ 
q_{\perp }%
\end{array}%
\right) 
=
\left( 
\begin{array}{ccc}
1 & 0 & 0 \\ 
0 & \cos \tau & \sin \tau \\ 
0 & \sin \tau & - \cos \tau%
\end{array}%
\right) 
\left( 
\begin{array}{c}
q_{1} \\ 
q_{2} \\ 
q_{3}%
\end{array}%
\right).   \label{qRot}
\end{equation}

	In the case of a rotating electric field Eq. (\ref{chi-e}) yields for the QED parameter
	\begin{equation}
	 \chi_e=\frac{a^2}{a_S}\frac{1+q_{1}^2+q_{\perp}^2}{\gamma_e}.
	\label{chi-eq}
	\end{equation}
	with $\gamma_e=\sqrt{1+q_{1}^2+q_{\perp}^2+q_{||}^2}$.
	
Substituting expressions (\ref{qRot}) into the equations of the electron motion (\ref{EquMot}), we obtain
	\begin{equation}
	 \frac{d q_{||}}{d \tau}+q_{\perp}=a
	 -\varepsilon_{rad}\, G_e(\chi_e) a^2 q_{||}\frac{q_{\perp}^2}{\gamma_e},
	\label{EquMotQpar}
	\end{equation}
	$$
	 \frac{d q_{\perp}}{d \tau}-q_{||}=
	 $$
	 \begin{equation}
	 -\varepsilon_{rad}\,G_e(\chi_e)\left[\gamma_e a+a^2 \frac{q_{\perp}}{\gamma_e}\left(1+ q_{\perp}^2 \right)\right].
	\label{EquMotQprp}
	\end{equation}
	Here and below we do not consider ultrarelativistic electron beam interaction with the laser radiation.
		We assume that the longitudinal component of the electron momentum, $q_1$, is much 
 less than the transverse momentum component $\sqrt{q_2^2+q_3^2}$, which corresponds to the case of 
 the laser pulse interacting with a near-critical density plasma, 
when $n\approx m_e \omega^2/4\pi e^2$ and $q_1^{(0)}\approx 1$, 
the change of the momentum component is negligible provided that the laser pulse duration 
is small enough.

Multiplying equation (\ref{EquMotQpar}) by $q_{||}/\gamma_e$ and equation (\ref{EquMotQprp}) 
by $q_{\perp}/\gamma_e$ and taking the sum we find
	\begin{equation}
	 \frac{d \gamma_e}{d \tau}=a \frac{q_{||}}{\gamma_e}
	 -\varepsilon_{rad} \,G_e(\chi_e) \left(a q_{\perp}+a^2 q_{\perp}^2 \right),
	\label{EquBal}
	\end{equation}
which shows how the electron acquires energy from the electromagnetic wave 
and loses it due to radiation friction.

\begin{figure*}[tbph]
\centering
\includegraphics[width=13.2cm,height=8.8cm]{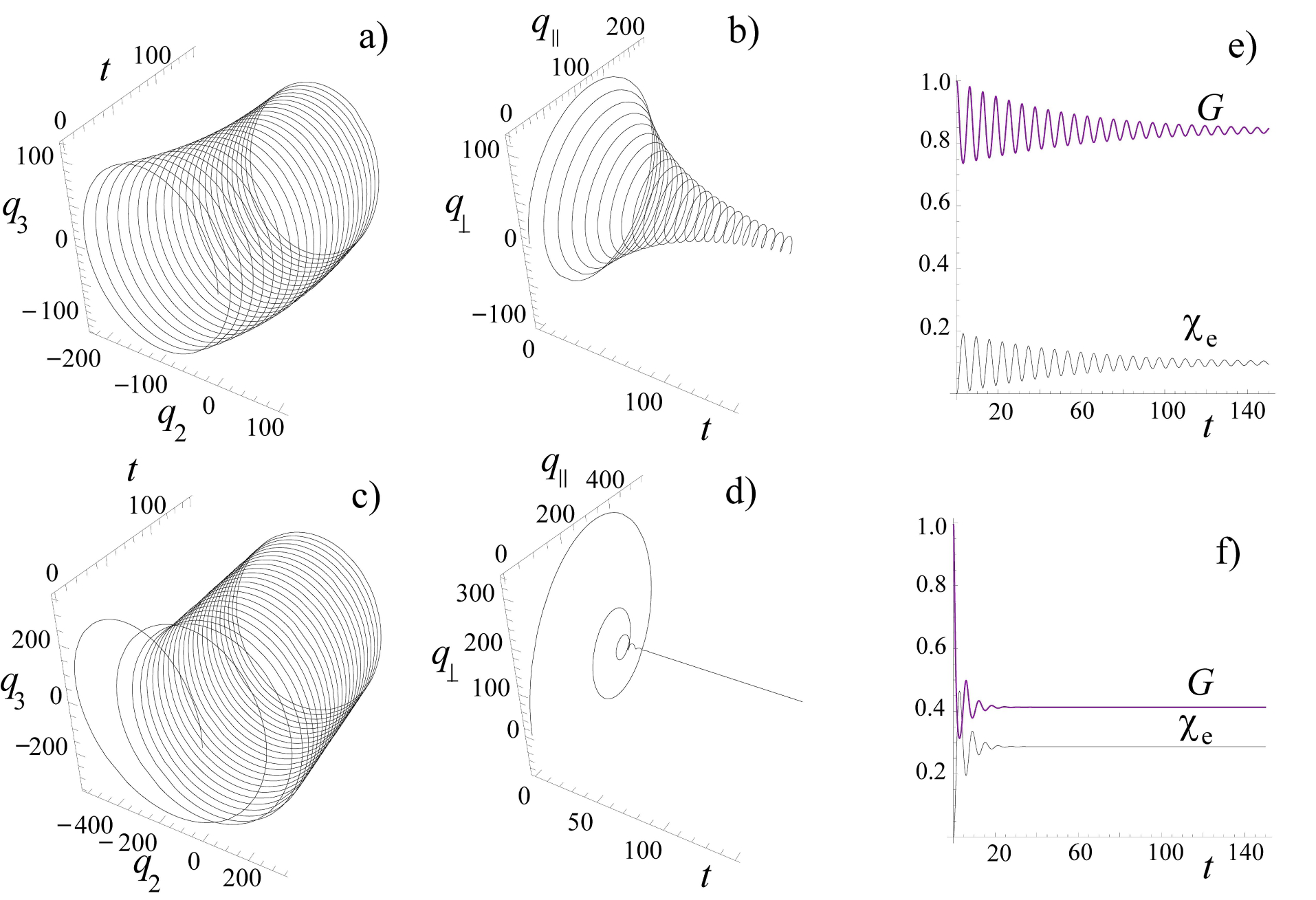}
\caption{Dependence of $q_{2}$ and $q_{3}$ (a,b), $q_{||}$ and $q_{\perp}$ (c,d) and $G_e$ and $\chi_e$ (e,f) 
on time for $\varepsilon_{rad}=10^{-8}$ and $a_S=4.1\times 10^5$.
(a,c) $a=0.25 \varepsilon_{rad}^{-1/3}$, (d,e) $a=0.75 \varepsilon_{rad}^{-1/3}$.
\label{FIG4}
}
\end{figure*}

Typical solutions of this system of equations are presented in Fig. \ref{FIG4}, 
where we show dependences of the $q_{2}$ and $q_{3}$ components of the electron momentum  with respect to time
time (l.h.s column, Fig. \ref{FIG4}a,c), and the  $q_{||}$ and $q_{\perp}$  components of the electron momentum 
 (central coloumn, Fig. \ref{FIG4}b,d). 
As we see, for $a \ll \varepsilon_{rad}^{-1/3}$ the electron oscillations in the ($q_{||}$,$q_{\perp}$)
 plane decay slowly while for the laser pulse amplitude values equal to or above 
 $\varepsilon_{rad}^{-1/3}$, 
 the electron oscillations in the rotating coordinate system 
decay during a time of the order of or less than the wave period. The dependence on time of the QED parameters $\chi_e(t)$ and $G(t)$  
is plotted in Fig. \ref{FIG4}e,f. We see that for $a=0.25 \varepsilon_{rad}^{-1/3}$, i.e. at the laser intensity of the order 
of $\approx 10^{21}$ W/cm$^2$ for $1 \mu$ m wavelength laser radiation the quantum reducing of the radiation friction force is negligibly small. 
Asymptotically at $t\to \infty$
we have $\chi_e\approx 0.1$ and $G_e^{1/3}\approx 0.95$. 
When $a=0.75 \varepsilon_{rad}^{-1/3}$ the quantum correction of the the radiation friction force becomes more significant with 
$\chi_e\approx 0.3$ and $G_e^{1/3}\approx 0.75$ in the limit $t\to \infty$.

Using the above obtained results  we can find  border lines between the domain where the dominant radiation friction regime takes place (this is the domain II in 
Fig. \ref{FIG3})
and the domain IV in Fig. \ref{FIG3}, where QED effects must be taken into account. They are given by different dependencies of the EM radiation intensity 
on the wave frequency. As was shown in Ref. \cite{LADvsLL}, in the limit $a>a_{rad}$ the transverse component of the electron momentum 
$q_{\perp}\approx (a \varepsilon_{rad})^{-1/2}$ is 
substantially less than the component parallel to the electric field  $q_{||}=(a/\varepsilon_{rad})^{1/4}$. Quantum effects become important, when 
the value of the QED parameter (\ref{chi-eq}), which becomes equal to $\chi_e\approx q_{||}/a_s$, 
approaches unity. This yields in the limit $\omega \ll \omega_1$ for the EM wave intensity 
	\begin{equation}
	 I_{R-Q}=\frac{m_e^4 c^5 e^2}{9 \pi \hbar^4}=5.6\times 10^{24}\frac{\rm W}{{\rm cm}^2}.
	\label{EquI-R-Q}
	\end{equation}

	The border line between domains III and IV for $\omega \gg \omega_1$ in Fig. \ref{FIG3} corresponds to the limit $\chi_e \gg 1$, when the asymptotic expression 
	for the form factor $G_e(\chi_2)$ is given by Eq. (\ref{G-chi-eInf}). From the equations of the electron motion (\ref{EquMotQpar}) and (\ref{EquMotQprp}) 
	we obtain, that the electron normalized enery depends on the EM field amplitude as $\gamma_e \approx (a^7/\varepsilon_{rad}^3 a_S^4)^{1/8}$. This yields for the 
	EM wave amplitude at the border line $a_{Q-R}\approx 96 a_S^{-4}\varepsilon_{rad}^{-3}\approx 324(\lambda^4_C/r_e^3 \lambda)$. Using this expression for 
	$a_{Q-R}$ we obtain a dependence of $I_{Q-R}(\omega)$ in the form
%
$$  I_{Q-R} \approx  $$
	\begin{equation}
	87 \frac{c^5 \hbar^8 \omega^4}{e^{14}}
	 =8\times 10^{21}\left(\frac{\omega}{\omega_1}\right)^4\frac{\rm W}{{\rm cm}^2}.
	\label{EquI-Q-R}
	\end{equation}

\begin{figure}[tbph]
\centering
 \includegraphics[width=7cm,height=5.5cm]{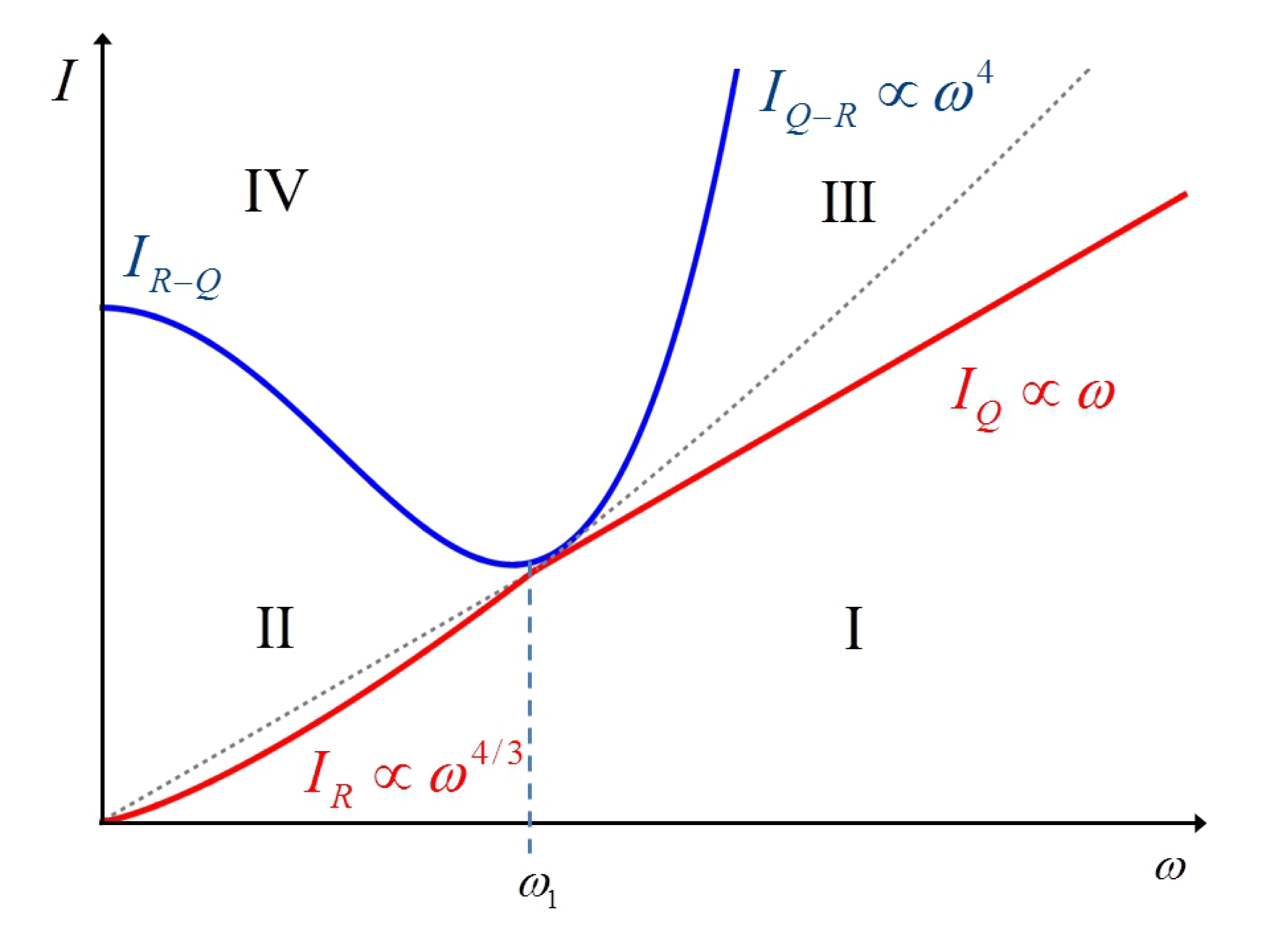}
 \caption{Curves $I_R(\omega)$ and $I_Q(\omega)$ subdivide the plane $I, \omega$ to four domains. I) Relativistic interaction of electron with the EM field 
when neither radiation friction nor QED effects are important. II) Electron -- EM wave interaction dominated by the radiation friction force effects. 
III) QED effects come into play with insignificant radiation friction force effects. IV) Both the QED and radiation friction force effects determine the 
radiating charged particle dynamics in the EM wave. \label{FIG5}}
\end{figure}

In Fig. \ref{FIG5} we plot the border curves $I_R, I_{R-Q},I_Q$ and $I_{Q-R}$
versus the EM wave frequency. These curves subdivide the plane $I, \omega$ to four domains. 
In the first domain, (I)  
 neither radiation friction nor QED effects are important for the relativistic interaction of an electron with the EM field. 
 In the second domain, 
 (II) the electron -- EM wave interaction is dominated by the radiation friction force effects. 
In the third domain, (III) the QED effects come into play with insignificant radiation friction force effects. 
In the high intensity limit, (IV) both the QED and radiation friction force effects determine the 
radiating charged particle dynamics in the EM wave. The dashed lines show the dependences $I_R(\omega)$ at $0<\omega<\omega_1$ and
$I_{Q}(\omega)$ at $\omega>\omega_1$.

%

The quantum nature of the photon emission process results in the recoil effect which causes the electron trajectory broadening 
(see review article \cite{TERNOV} and 
literature cited in). Following this paper we can write equation describing dependence on time of the mean square of the electron trajectory deflection
\begin{equation}
\frac{d \Delta r^2}{dt}=\frac{55}{48  \sqrt{3}}\frac{r_e c \lambda_C}{\lambda}\gamma_e^5.
\label{eq:Dr2}
\end{equation}
A condition of the electron deflection during the laser wave period $2\pi/\omega$ being of the order of the laser wavelength, 
$\Delta r \approx \lambda$,  
shows that the quantum diffusion 
becomes important at the electron quiver energy corresponding to 
\begin{equation}
\gamma_D=\left(\frac{24 \sqrt{3}}{55 \pi}\frac{m_e^2 c^5}{\hbar \omega^2 \hbar e^2}\right)^{1/5}
=\frac{2.4}{\alpha^2} \left(\frac{\omega_1}{\omega}\right)^{2/5},
\label{eq:gDr2}
\end{equation}
i.e. above approximately 25 GeV. Corresponding intensity is
\begin{equation}
I_D=7.2 \times 10^{26}\left(\frac{\omega_1}{\omega}\right)^{4/5} \frac{{\rm W}}{{\rm cm^2}}.
\label{eq:ID}
\end{equation}

 \section{Integral Scattering Cross Section}
 
     According to Eq. (\ref{EquBal}), the energy flux reemitted by the electron is equal to
	 $$ e\left({\bf v}\cdot{\bf E}\right) \approx  $$
	\begin{equation} 
	 \varepsilon_{rad} G_e(\chi_e) m_e c^2 \omega \gamma_e \left(a q_{\perp}+a^2 q_{\perp}^2 \right).
\end{equation} 
This expression is a condition of the balance between the rate of energy acquired by the electron from the EM field and the rate of the radiation losses.
	 The integral scattering cross section by definition \cite{LL-TP} equals the ratio of the 
	 reemitted energy flux to the Poynting vector amplitude, $cE^2/4\pi$. This yields an 
	 expression for the cross section dependence on the electron momentum
	 \begin{equation}
	 \sigma=\sigma_T G_e (\chi_e)\left(q_{\perp}^2+\frac{q_{\perp}}{a}\right).
	\label{EquSigma1}
	\end{equation}
	Here $\sigma_T$ is the Thomson scattering cross section, $\sigma_T=8\pi r_e^2/3=6.65 \times 10^{-25}$cm$^2$. 
Expression (\ref{EquSigma1}) can also be approximated by 
	\begin{equation}
	 \sigma=\sigma_T \left(\frac{\gamma_e^2 G_e(\chi_e)}{1+\varepsilon_{rad}^2 G_e^2(\chi_e) \gamma_e^6}\right).
	\label{EquSigma2}
	\end{equation}

\begin{figure}[tbph]
\centering
   \includegraphics[width=7cm,height=7cm]{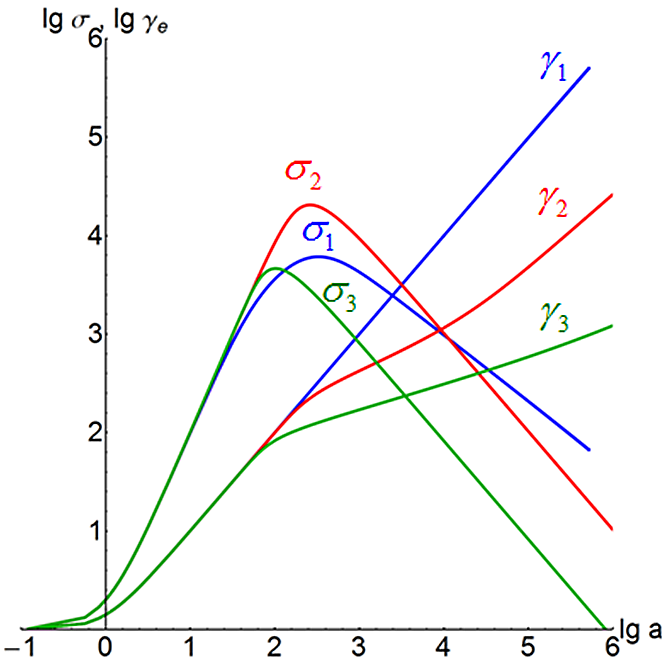}
	 \caption{ Dependences of ${\rm lg}(\sigma/\sigma_T)$ and of ${\rm lg}(\gamma_e)$ on ${\rm lg}(a)$.  
  1) $\omega=\omega_1/12.5$, 
  2)  $\omega=\omega_1$, 
 3)  $\omega=12.5 \omega_1$.
  \label{FIG6}}
 \end{figure}

In the range of the wave amplitudes of $1\ll a \ll (\varepsilon_{rad} G_e)^{-1/3}$, we have $q_{\perp} \gg q_{||}$ with $q_{\perp}\approx a$. 
The integral 
scattering cross section grows as 
\begin{equation}
\sigma\approx \sigma_T \left(1+a^2 \right).
\end{equation}
For $\chi_e \ll 1$, i.e. 
\begin{equation}
1\ll \varepsilon^{-1/3}_{rad}\ll a \ll \varepsilon_{rad} a_S^2,
 \end{equation}
it reaches a maximum of $\sigma\approx 0.53 \sigma_T \varepsilon^{-2/3}_{rad}$ at $a=1.1 \varepsilon^{-1/3}_{rad}$.
These dependences correspond to the domains (I) and (III) in Fig. \ref{FIG6}, where the radiation friction effects are weak.
Then for $a\gg \varepsilon^{-1/3}_{rad}$ 
the scattering cross section decreases according 
to 
\begin{equation}
\sigma\approx \sigma_T/a\varepsilon_{rad},
\end{equation}
	as seen in Fig. \ref{FIG6} as a consequence of the fact that the maximal
power reemited by the electron cannot exceed $eEc$. This gives a constraint on
the scattering cross section: $4 \pi e/ E$.  In the limit $\varepsilon^{-1/3}_{rad}\ll a \ll \varepsilon_{rad} a_S^2$
the electron momentum components are $q_{\perp} \approx (a\varepsilon_{rad})^{-1}$ and $q_{||}\approx (a/\varepsilon_{rad})^{1/4}$ 
with $q_{\perp} \ll q_{||}$ \cite{LADvsLL}, which corresponds to domain (II) in Fig. \ref{FIG6}, where the radiation friction is dominant. 
 
 Dependences of the scattering cross section normalized by $\sigma_T$ and the electron energy $\gamma_e$ 
 on the laser pulse amplitude $a$ are shown in Fig. \ref{FIG6}. The parameters 
 $\varepsilon_{rad}=1.75\times 10^{-8}$ and $a_S=2.8\times 10^{5}$ (curves 2) correspond to the EM frequency equal to $\omega_1$ given by Eq. (\ref{eq:om1}). 
 For $\omega=\omega_1/12.5$ (curves 1) the EM wave length is about $\lambda=10 \mu$m, which is typical for the CO$_2$ laser wavelength range. 
 The case 3 with $\omega=12.5 \omega_1$ corresponds to 
  the parameters in the domain (IV) in Fig. \ref{FIG6} with the radiation friction lowered by quantum effects.

	 \section{Computer Simulations of the Radiation Friction and QED Effects on the 
	Laser Pulse Interaction with Inhomogeneous Plasma }
	
	The electron quivering in the laser field emits photons whose energy is proportional to 
	 the cube of the electron Lorentz factor: $\hbar \omega_{\gamma}\approx 0.3 \hbar \omega \gamma_e^3$. 
	 For $a\ll \varepsilon_{rad}^{-1/3}$ a typical value of the photon frequency 
	 is proportional to $a^3$. In the limit of high laser intensity $a\gg \varepsilon_{rad}^{-1/3}$ the frequency scales as 
	 $\omega_{\gamma}=\omega(a/\varepsilon_{rad})^{3/4}$. For multi-petawatt laser radiation the emitted photon energy is 
	 in the gamma ray energy range.
	 
\begin{figure}[tbph]
\centering
\includegraphics[width=6cm,height=4.25cm]{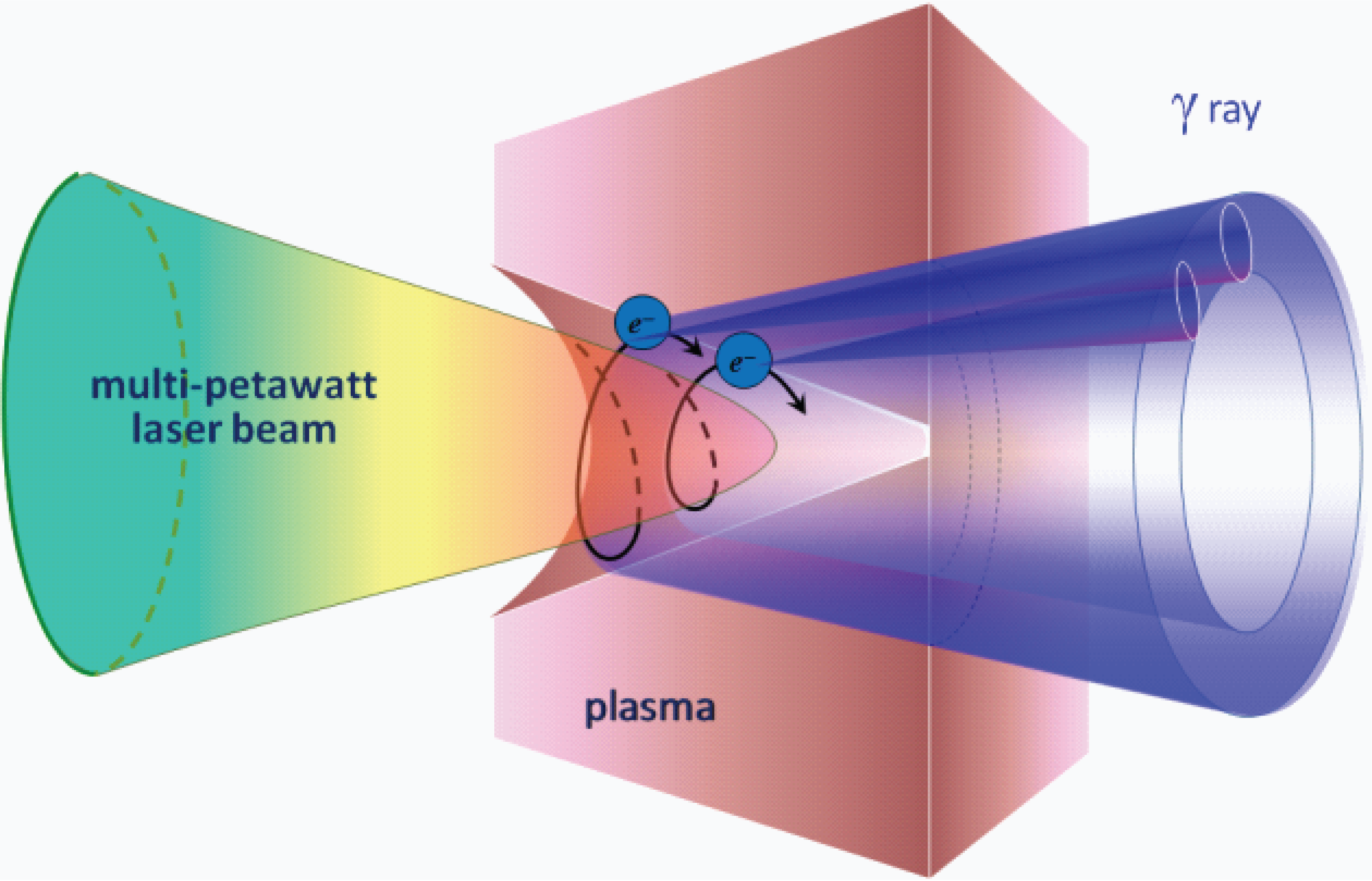}
\caption{The concept  of  high power gamma-ray flash generation in the laser-matter interaction.
}
\label{FIG7}
\end{figure}

	 The gamma-ray pulse energy, duration and divergence are determined by 
the laser pulse amplitude and by the plasma target density scale length. 
We analyse the density scale length effect on the parameters of the emited gamma-flash. Fig. \ref{FIG7}
illustrates the concept  of the high power gamma-ray flash generation in the laser-matter interaction.

     During interaction of super-high-power laser light with matter the laser pulse 
	 is subject to various instabilities. Among them the most important is the relativistic 
	 self-focusing resulting in the laser pulse modulation and channeling. It leads to the increase
	 of the laser pulse amplitude and to the decrease of the electron density in the interaction 
	 region, which change the laser energy depletion length and the parameters of the gamma-rays emitted. 
	 Thorough studying of these effects and of the effects of the plasma inhomogeneity require 
	 computer simulations.

     We performed studies of the laser pulse interaction with high density targets 
	 using a two-dimensional (2D) particle-in-cell (PIC) code 
	 where the radiation friction 
	 force has been incorporated in the Landau-Lifshitz form as has also been done in Refs. \cite{NMN}.
	 In addition, the QED radiation friction weakening is taken into account by multiplying the radiation force 
	 by a form-factor $G_e(\chi_e))$ given by Eq. (\ref{G-chi-eappr}). We note that results of 
	 detailed computer simulations of the high power gamma ray flash emission by a multi-petawatt 
     pulse laser have been presented in Ref. \cite{NakaKo}, where the QED contribution was assumed to be negligebly small.	 
	 In that paper the plasma target inhomogeneity has been suggested for optimization of the EM wave energy 
	 convertion to the gamma ray energy flash.
	 
     In the simulations, the laser pulse has the power  ${\cal P}_{\rm las}$ is equal to $100$PW 
	 	 with the pulse duration of ${\cal T}_{\rm las}=30\,$fs. 
	 
	 In the tailored plasma target the density changes 
	 exponentially, $n\propto \exp (x/L)$, with the plasma inhomogeneity scale length, $L$, 
	 from $0.1\,n{\rm cr}$ to $350\,n{\rm cr}$ in the interval  $ \approx 20\,{\rm \mu m}$,
	 and then becomes constant having a thickness of $10\,{\rm \mu m}$. 
	 	 
	 The simulation box has a 
	 width equal to $80\,{\rm \mu m}$ and a length varying from $50\,{\rm \mu m}$ to $210\,{\rm \mu m}$. 
	 The mesh has a spatial resolution of  $\Delta x=\Delta y$ varying from $1/40\,{\rm \mu m}$ 
	 to $1/200\,{\rm \mu m}$ with a temporal resolution of $\Delta t=0.0025\,$fs. 
	 The plasma is comprised of electrons and ions with a mass to charge ratio corresponding to $A/Z=2$,
 corresponding to fully ionized Carbon. The average number of quasiparticles is about $4\times 10^8$.

	\begin{figure*}[tbph]
\centering
    \includegraphics[
	width=12cm,height=6cm
	]{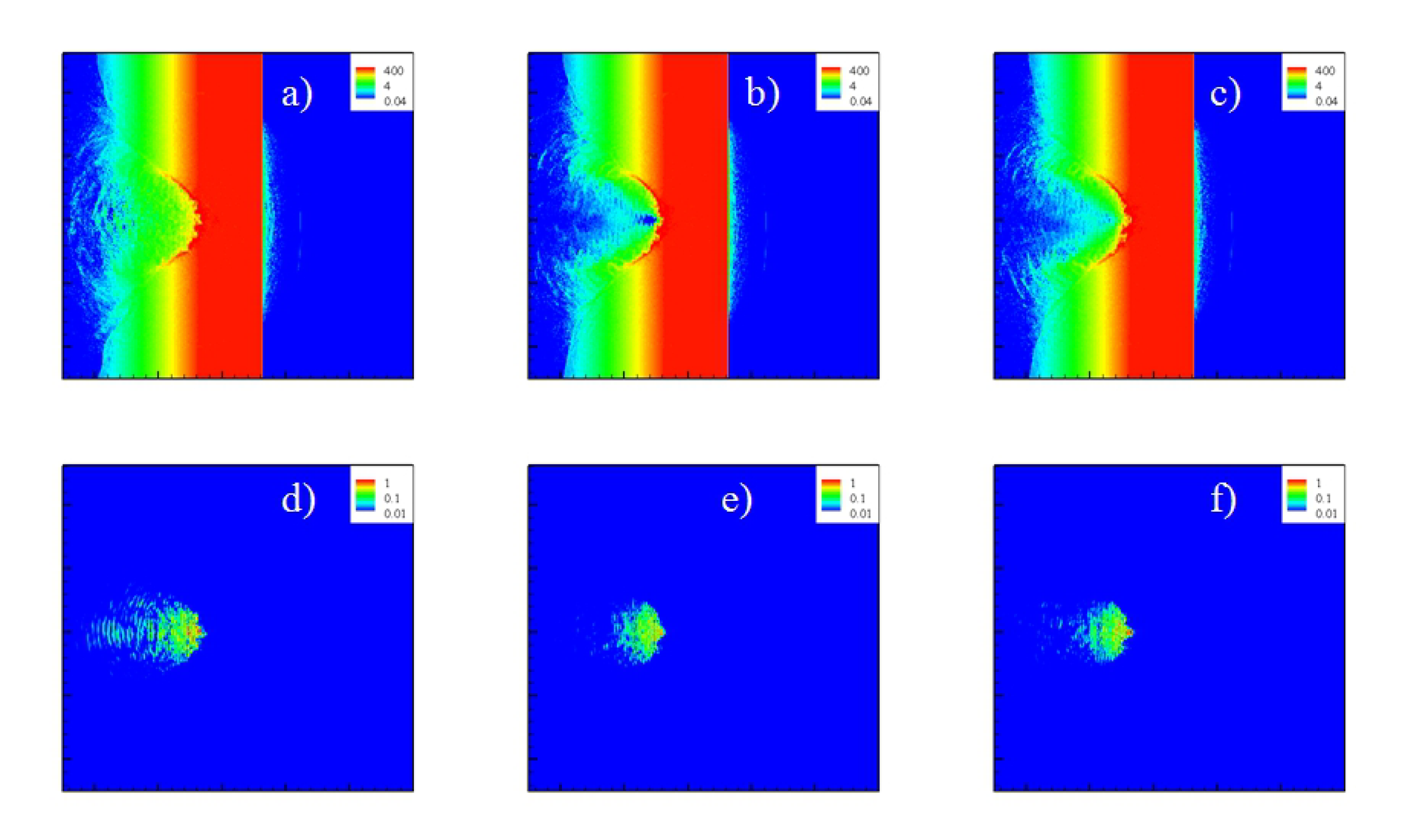}
	\caption{ Computer simulation of a 100 PW laser pulse interaction with a tailored target. 
	a) Time evolution of the electron ${\cal E}_e$, ion ${\cal E}_i$ and total energy ${\cal E}_{\rm tot}$ with no radiation friction force and QED effects.
	b) ${\cal E}_e$, ion ${\cal E}_i$, gamma-ray ${\cal E}_{\gamma}$ and total energy ${\cal E}_{\rm tot}$ with the radiation friction force but without QED effects.  
	c) ${\cal E}_e$, ion ${\cal E}_i$, gamma-ray ${\cal E}_{\gamma}$ and total energy ${\cal E}_{\rm tot}$ with the radiation friction force and with 
the	QED effects. 
	d) The gamma ray pulse power ${\cal P}_{\gamma}$ and energy ${\cal E}_{\gamma}$ vs time with the radiation friction force but without 
QED effects. 
e) ${\cal P}_{\gamma}$ and  ${\cal E}_{\gamma}$ vs time with the radiation friction force and QED effects. 
Time is measured in fs. Units of the energy and power are $J$ and $PW$, respectively.
	\label{FIG8}	}
	\end{figure*}
	 
	 Simulation results for the parameters 
	 of interest are shown in Figs. \ref{FIG7} and \ref{FIG8}. 
	 The 100 PW laser pulse with the normalized amplitude $a=474$ interacts with the plasma target, whose 
	 density inhomogeneity is characterized by a scale length equal to $L=2.5\,{\rm \mu m}$. 
	 
	 Fig. \ref{FIG8} a) shows the dependence on time of the electron ${\cal E}_e$, ion ${\cal E}_i$ and total energy ${\cal E}_{\rm tot}$ 
	 when the radiation friction force and the QED effects are neglected. We see that the laser pulse energy at first is converted 
	 to the electron energy and then to the ion energy. In Fig. \ref{FIG8} b) we plot the same dependences of the electron, 
	 ion and total energy and 
     the energy of emitted gamma-rays in time for the case with the radiation friction force taken into account but without QED effects. 
	 In this case 
	 a major part of the laser pulse energy is transferred to the gamma-ray flash. Fig. \ref{FIG8} d) presents 
	 the gamma ray pulse power ${\cal P}_{\gamma}$ and energy ${\cal E}_{\gamma}$ versus time in this case. The maximal gamma ray pulse power is equal to 
	 45 PW. If the radiation friction force and the	QED effects are incorporated into the model the conversion of the laser pulse energy to the gamma ray 
	 radiation becomes lower as it is seen in Figs. \ref{FIG7} c) and e). The maximum gamma ray pulse power becomes equal to 
	 40 PW.
	 
\begin{figure*}[tbph]
\centering
    \includegraphics[
	width=12cm,height=6cm
	]{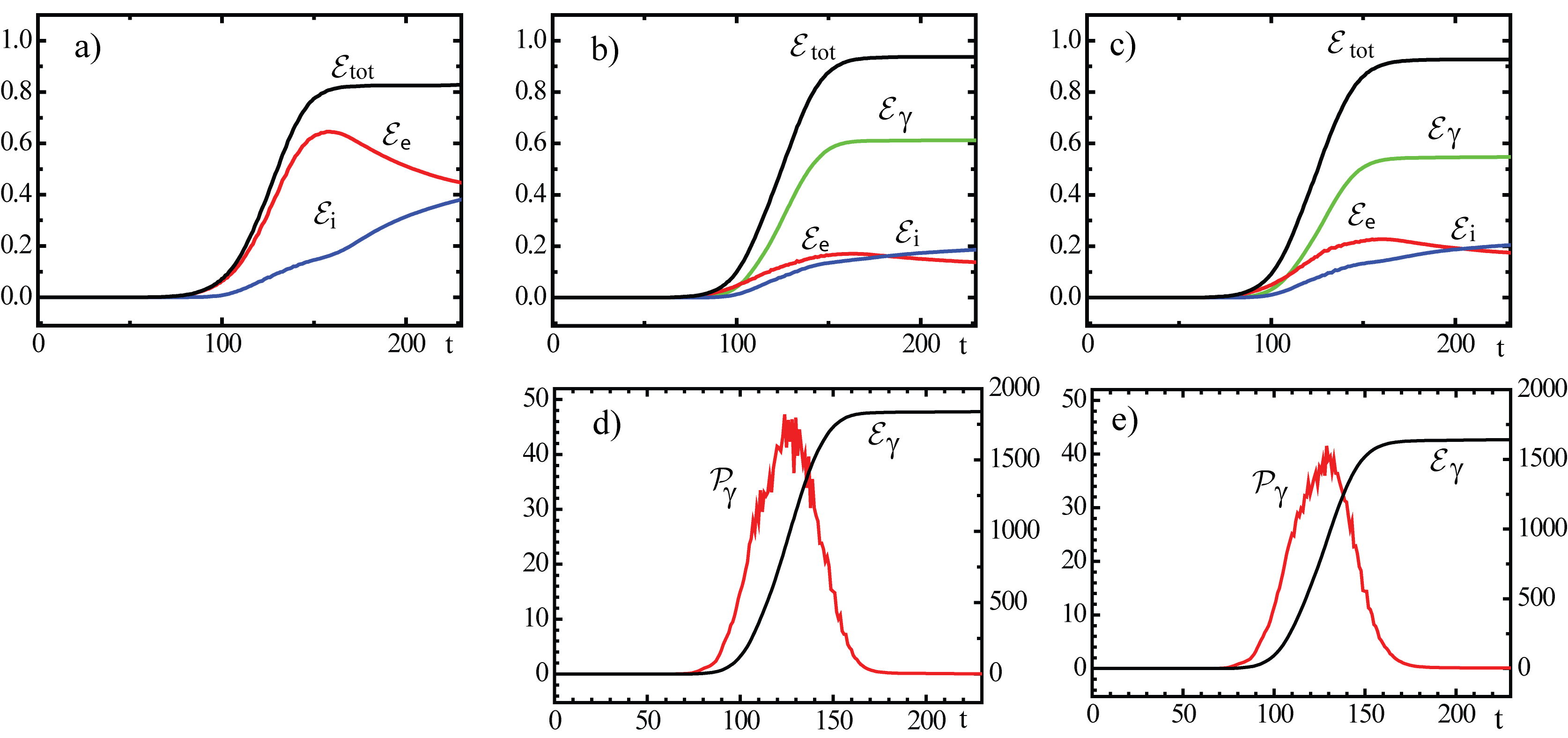}
	\caption{ Computer simulation of the hole boring by a 100 PW laser in the tailored target. a,b,c) The electron density 
	 distribution and d,e,f) and the laser electric field in the $(x,y)$ plane at $t=140$fs. For a) and d) no radiation friction force and QED effects
	 are taken into account. b) and e) show the hole boring, when the radiation friction force is present but without the QED effects.  
	e) and f) correspond to the case  of the nonvanishing radiation friction force and QED effects. 
	\label{FIG9}}
		\end{figure*}

	 Fig. \ref{FIG9} illustrates the hole boring by the laser pulse in the tailored plasma targets 
	 when neither the radiation friction force nor QED effects are taken into account (see \ref{FIG9} a), where the electron density 
	 distribution in the $(x,y)$ plane is shown at $t=140$fs and d) with the laser electric field distribution presented for the same time), when 
	 the radiation friction force is present but QED effects are neglected (\ref{FIG9} b) and e)), and when 
	 the radiation friction force and QED effects are incorporated into the description (\ref{FIG9} c) and f)). 
	 The radiation friction makes the hole walls less modulated and the laser pulse less filamented compared to the case when the radiation 
	 force effects are neglected. The QED corrections lower the radiation force resulting in a slightly more modulated laser pulse and 
	 hole wall.
	 
\begin{figure}[tbph]
\centering
    \includegraphics[
	width=6cm,height=4.8cm
	]{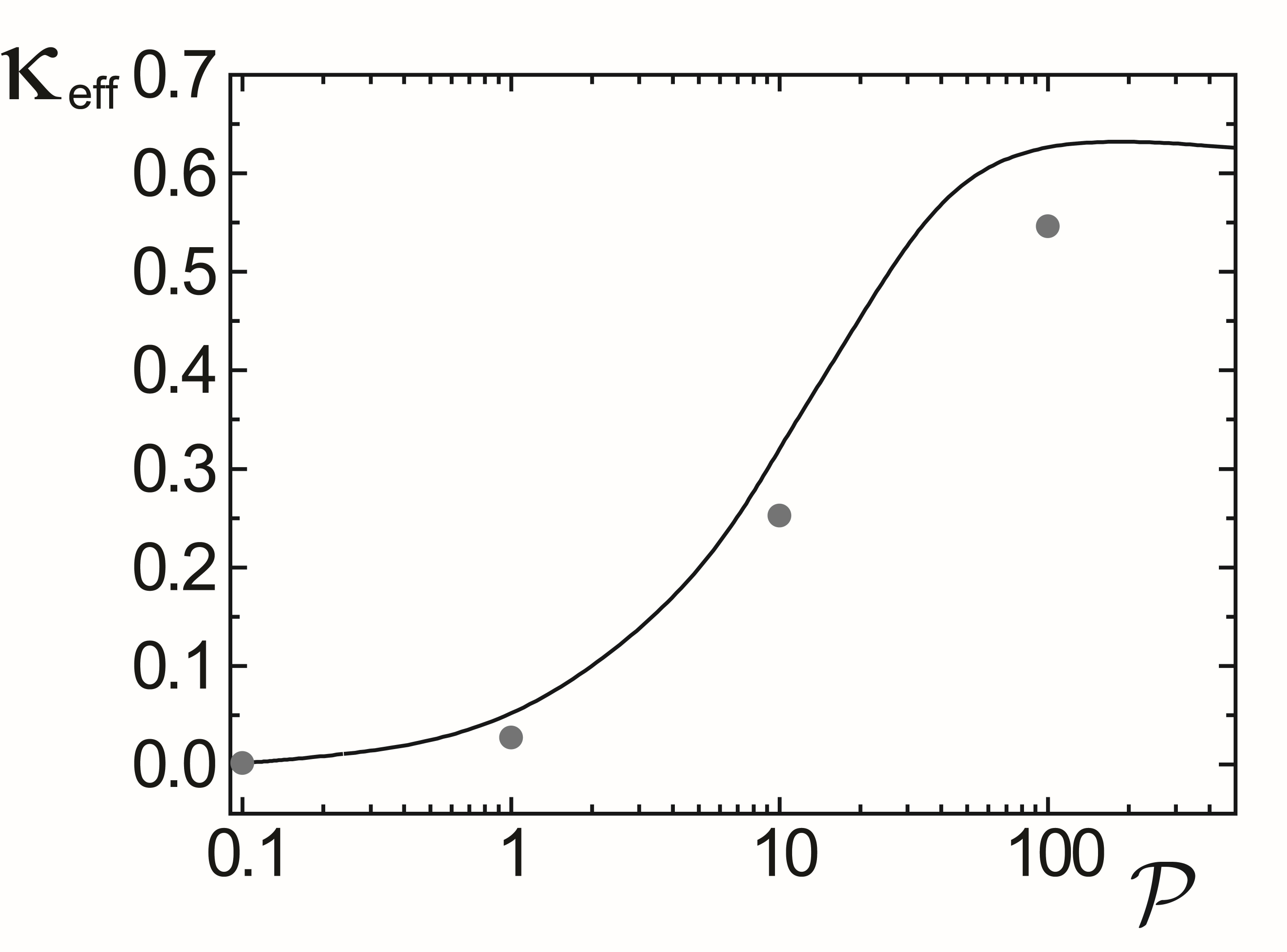}
		\caption{Comparison of the coversion efficiency of the laser energy to the energy of the gamma-rays 
	for a 30 fs laser pulse interacting with the tailored target when the quantum correction of the radiation friction is taken into account (dots) and 
	when it is neglected (solid curve).
	\label{FIG10}}
		\end{figure}

		The efficiency of the laser energy conversion to the energy of the gamma-ray flash, $\varkappa_{eff}$, depends on the laser pulse power. 
		In Fig. \ref{FIG10} we compare the coversion efficiency with and without the quantum correction of the radiation friction for 30 fs laser 
		pulses with the power in the range from 0.1 PW to 100 PW. As we see in the relatively low power limit the quantum effects are negligebly small.
		At the laser power equal to 10 PW the conversion efficiency is 0.25 and 0.3, and for the 100 PW laser pulse we have 0.55 and 0.65, respectively.

	\section{Spectrum of the radiation emitted by an ensemble of ultrarelativistic electrons}
	
	The frequency spectrum of a relativistic electron rotating along the trajectory with the radius of curvature, $R$,
is given by the expression, 
$$\frac{dI_{\gamma}(\omega_{\gamma},\cal E)}{d\omega_{\gamma}}
		= $$
\begin{equation}
\frac{\sqrt{3}}{2\pi} \frac{e^2}{c} \frac{\cal E}{m_e c^2}u(\omega_{\gamma},{\cal E})
		\int\limits_{u(\omega_{\gamma},{\cal E})}^{\infty}
		K_{5/3}(x)d x.
\label{eq-synchro}
\end{equation}
which is well known in the classical electrodynamics \cite{LL-TP}. Here $K_{\nu}(z)$ is the modified Bessel function \cite{AS}.
The electron energy is assumed to be ${\cal E}\gg m_ec^2$,
 $m_e$ and $c$ are the electron mass and the speed of light in vacuum, respectively. 
The function of the electron energy, $u(\omega_{\gamma},{\cal E})=\omega_{\gamma}/\omega_c$, 
is the ratio of the ratio of the emitted photon frequency $\omega_{\gamma}$ to 
the critical frequency equal to 
	\begin{equation}
	\omega_c=\frac{3 c}{2 R}\left( \frac{\cal E}{m_e c^2}\right)^3.
	\label{omc}
	\end{equation}
In the limit $u(\omega_{\gamma},{\cal E})\ll 1$ expression (\ref{eq-synchro}) yields $dI/d\omega_{\gamma}\approx (e^2/c)(\omega_{\gamma} R/c)^{1/3}$,
while for $u(\omega_{\gamma},{\cal E})\gg 1$ we have 
	\begin{equation}
\frac{dI}{d\omega_{\gamma}}\approx \sqrt{\frac{3 \pi}{2}}\frac{e^2}{c}\frac{{\cal E}}{m_ec^2} u(\omega_{\gamma},{\cal E})^{2}
e^{-u(\omega_{\gamma},{\cal E})}.
\end{equation}

	Electrons rotating along the circles with the same radius emit electromagnetic radiation with the same frequency spectrum independently 
	of the particular radiation mechanism. 
	
	In the case of synchrotron radiation when the ultrarelativistic electron with the energy ${\cal E}$ 
	rotates in a homogeneous magnetic field the Larmor radius 
	in the limit ${\cal E}\gg m_ec^2$ is equal to $r_B={\cal E}/eB$. 
	The characteristic photon frequency of the synchrotron radiation according to Eq. (\ref{omc}) is given by
	$\omega_c=(3 /2)\omega_B({\cal E}/{m_e c^2})^2$, where $\omega_B=eB/m_e c$ is the Larmor frequency.
	
	The characteristic energy of a photon emitted via nonlinear Thomson scattering of the circularly polarized electromagnetic wave scales 
with the electron quiver energy as 
$\omega_c=(3 /2)\omega_0({\cal E}/{m_e c^2})^3$,
where $\omega_0$ is the laser frequency. The energy of the 
electron quivering in plasma under the action of an electromagnetic wave with an amplitude 
of $a=eE/m_e \omega c \gg 1$ is of the order of ${\cal E}=a m_e c^2$. 
For a laser frequency of the order of $10^{15}$s$^{-1}$ the 
emitted photon energy 
is in the gamma-ray range if $a>10^{2}$ which corresponds to a laser intensity higher 
than $10^{22}$W/cm$^2$. 

Computer simulations of a petawatt power short laser pulse interaction with a near-critical plasma  \cite{NakaKo} show that
	the electron energy spectra in the interval ${\cal E}<{\cal E}_m$
	can be approximated by a power-law dependence. 
	For the energy ${\cal E}<{\cal E}_m$ the spectrum has exponential form.  Under the conditions of the simulations presented 
	in Ref. \cite{NakaKo} the energy, ${\cal E}_m$, is of the order of 50 MeV.
	The electron energy distribution can be described by the dependence
 \begin{equation}
		\frac{dN ({\cal E})}{d{\cal E}}=K {\cal E}^{-\kappa}\exp\left(-\frac{\cal E}{{\cal E}_m}\right).
		\label{kappa}
	\end{equation}
	For the parameters corresponding to the results presented in \cite{NakaKo}  the power
	index equals $\kappa \approx 0.8$. 
		
	The averaged spectrum of the emitted photons is given by the integral
	\begin{equation}
		J(\omega_{\gamma})
		=\int\limits_{0}^{\infty}
		\frac{dI_{\gamma}(\omega_{\gamma},{\cal E})}{d\omega_{\gamma}}
		\frac{dN ({\cal E})}{d{\cal E}} d {\cal E},
	\label{synch-aver}
	\end{equation}
	where the functions $dI_{\gamma}/d\omega_{\gamma}$ and $dN ({\cal E})/d{\cal E}$ are determined by Eqs. (\ref{eq-synchro}) and (\ref{kappa}).

	We analyze the spectrum of the radiation emitted in the process of nonlinear Thomson scattering of a circularly polarized electromagnetic wave, when 
the function $u(\omega_{\gamma},{\cal E})$ is
\begin{equation}
		u_C(\omega_{\gamma},{\cal E})=\frac{2}{3}\frac{\omega_{\gamma}}{\omega_0}\left(\frac{m_e c^2}{\cal E}\right)^3.
	\label{u-NTS}
	\end{equation}

		Substituting the functions (\ref{eq-synchro}) and (\ref{kappa}) to (\ref{synch-aver}) and performing integration over ${\cal E}$
		 we obtain the averaged spectrum of the emitted photons
		\begin{equation}
	J_C(\omega_{\gamma})
		=Q_C\,
		\omega_{\gamma}^{\frac{2-\kappa}{3}}
		F_C\left(\epsilon_m,\kappa\right),
	\end{equation}
	where
	\begin{equation}
	Q_C=\frac{\sqrt{3} K}{2\pi} 
		\frac{e^2}{c} 
		\left( \frac{2}{3 \omega_0}\right)^{\frac{2-\kappa}{3}} {(m_ec^2)^{(1-\kappa)}},
	\end{equation}
	\begin{equation}
	\epsilon_m=\frac{{\cal E}_m}{m_e c^2}\left( \frac{3 \omega_0}{2\omega_{\gamma}}\right)^{\frac{1}{3}}
	\end{equation}
	and
	$$ F_C\left(\epsilon_m,\kappa\right)= $$
	\begin{equation}
			\epsilon_m^{-(1+\kappa)}\int\limits_{0}^{\infty} 
			K_{5/3}(x)\Gamma\left(-1-\kappa,\frac{1}{x^{\frac{1}{3}}\epsilon_m}\right)dx.
			\label{Fepkapp}
	\end{equation}
	Here $\Gamma(a,z)$ is the incomplete gamma function \cite{AS}. The function $F_C\left(\epsilon_m,\kappa\right)$ 
	can be expressed in terms of hypergeometric functions.
	
	In the low frequency range, $\omega_{\gamma}\ll (3/2) \omega_0 ({\cal E}_m/m_e c^2)^3$, 
	using the asymptotic representation of the incomplete gamma function at $x^{-\frac{1}{3}}\epsilon_m^{-1}\to 0$,
	$$ \Gamma\left(-1-\kappa,\frac{1}{x^{\frac{1}{3}}\epsilon_m}\right)=  $$
	\begin{equation}
			\Gamma(-1-\kappa)+\epsilon_m^{-(1+\kappa)}\frac{x^{\frac{1+\kappa}{3}}}{1+\kappa}
			-\epsilon_m^{-\kappa}\frac{x^{\frac{\kappa}{3}}}{\kappa}+...\, ,
	\end{equation}
	we obtain 
	\begin{equation}
			F_C\left(\infty,\kappa\right)=\frac{2^{\frac{\kappa-2}{3}}}{1+\kappa}\Gamma\left(\frac{\kappa-1}{6}\right)\Gamma\left(\frac{\kappa+9}{6}\right),
	\end{equation}
	where $\Gamma(x)$ is the gamma function \cite{AS}. This expression is valid provided $\kappa>1$. 
	We see that the spectrum has the power form, 
	\begin{equation}
	J_C(\omega_{\gamma})\sim \omega_{\gamma}^{\frac{2-\kappa}{3}}. 
	\end{equation}

If the power index is greater than 2, $\kappa>2$, the dependence $J_C(\omega_{\gamma})$ is monotonically decreasing as 
$\omega_{\gamma}/\omega_0 \to \infty$. For $1<\kappa<2$ the function $J_C(\omega_{\gamma})$ grows with $\omega_{\gamma}$. 
In this case as well, if the power index 
is less than unity, $\kappa<1$,  we should take into account that the electron energy spectrum is truncated at the
high energy end.

Analyzing the asymptotic behaviour of the function $F_C\left(\epsilon_m,\kappa\right)$ 
given by Eq. (\ref{Fepkapp}) as $\epsilon_m\to \infty$ for $\kappa<1$ we find 
the low frequency scaling ($\omega_{\gamma}\ll (3/2) \omega_0 ({\cal E}_m/m_e c^2)^3$, $\kappa<1$)
$$ J_C(\omega_{\gamma})\approx$$
\begin{equation}
\frac{\sqrt{3}e^2}{\pi c} 
\left(\frac{{\cal E}_m}{m_e c^2} \right)^{(1-\kappa)}
\Gamma\left(\frac{2}{3}\right)
\Gamma\left(1-\kappa\right)
\left( \frac{ \omega_{\gamma}}{3 \omega_0}\right)^{1/3}.
\label{eq:Ju-s0-kg1-F3}
\end{equation}

Considering the limit $\epsilon_m \to 0$, which corresponds to the high frequency 
range, $\omega_{\gamma}\gg (3/2) \omega_0 ({\cal E}_m/m_e c^2)^3$, 
we obtain
$$ F_C\left(\epsilon_m,\kappa\right) \approx $$
$$ \sqrt{\frac{\pi}{2}}\epsilon_m^{-\kappa}
			\int\limits_{0}^{\infty}dx x^{\frac{1+2 \kappa}{6}} 
			\exp {\left(-x-\frac{1}{x^{1/3}\epsilon_m}\right)}  $$
\begin{equation}
			 \propto \exp{\left(-\frac{4}{3^{3/4}\epsilon_m^{3/4}}\right)},
				\end{equation}
i. e., the spectrum decreases exponentially with photon frequency,
\begin{equation}
	J_C(\omega_{\gamma})
	\sim 
	{\rm exp}
	\left[
	-4\left(
	\frac{2 \omega_{\gamma} m_e^3 c^6}{51 \omega_0 {\cal E}_m^3}
	\right)^{\frac{1}{4}}
	\right]. 
	\end{equation}
 As a result we see that the frequency spectrum has a maximum at 
\begin{equation}
	\omega_{\gamma, m}\approx 14.2\times\omega_0 \left(\frac{{\cal E}_m}{m_e c^2}\right)^3.
	\label{ommax} 
	\end{equation}
	For the parameters of the simulations presented in \cite{NakaKo} the 
	characteristic energy ${\cal E}_m$ is of the order of 38 MeV. 
	This yields for the photon energy, $\varepsilon_{\gamma,m}=\hbar \omega_{\gamma, m}$, corresponding 
	to the spectrum maximum $\varepsilon_{\gamma,m}\approx 6$ MeV.

\section{Conclusion}

With Kilo-Joule lasers high field science will enter novel regimes of electromagnetic radiation interaction 
with matter when radiation friction force effects result in  the high efficiency conversion of the energy of the EM wave into the
energy of hard EM radiation in the form of ultra short high power gamma ray flashes. 
The energy spectrum of the gamma-ray flash emitted by the relativistic electrons has a typical form with a maximum which 
dependence on the parameters is given by Eq. (\ref{ommax}).

\section*{Acknowledgments}
We thank E. Esarey, T. Grismayer, O. Klimo, J. Limpouch, N. B. Narozhny, N. N. Rosanov, and L. Silva for
discussions. This work was supported by the Ministry of Education,
Science, Sports and Culture of Japan, grant - Kiban(C)25400540.


\end{document}